\documentstyle[12pt,epsfig]{article} 

\def\bb{b \bar{b}} 
\def\cc{c \bar{c}}

\def\cO{{\cal O}} 
\def\cF{{\cal F}}

\begin{document}  
\vspace*{-2cm}  
\renewcommand{\thefootnote}{\fnsymbol{footnote}}  
\begin{flushright}  
hep-ph/9907238\\
DTP/99/70\\  
%DRAFT: \today\\  
July 1999\\  
\end{flushright}  
%\vskip 65pt  
\vskip 45pt  
\begin{center}  
{\Large \bf Higgs Boson Production at the Compton Collider}\\
\vspace{1.2cm} 
{\bf  
Michael Melles${}^1$\footnote{Michael.Melles@durham.ac.uk} and  
W.~James~Stirling${}^{1,2}$\footnote{W.J.Stirling@durham.ac.uk}  
}\\  
\vspace{10pt}  
{\sf 1) Department of Physics, University of Durham,  
Durham DH1 3LE, U.K.\\  
  
2) Department of Mathematical Sciences, University of Durham,  
Durham DH1 3LE, U.K.}  

\vspace{0.6cm} 

and

\vspace{0.6cm} 

{\bf Valery A. Khoze${}^{1,3}$\footnote{Valery.Khoze@cern.ch}}\\ 

\vspace{10pt}  

{\sf 3) INFN-LNF, P.O.Box 13, I-00044, Frascati (Roma), Italy}

\vspace{20pt}  
\begin{abstract}
The high precision determination of the partial width $\Gamma ( H \longrightarrow
\gamma \gamma )$ of an intermediate mass Higgs boson is among the most important 
measurements at a future photon--photon collider. Recently it was shown that
large  non-Sudakov as well as Sudakov double logarithmic corrections
can be summed to all orders in
the background process $\gamma \gamma \longrightarrow q \overline{q}$, $q=\{b,c\}$,
from
an initially polarized $J_z=0$ state.
In addition, running coupling corrections were included exactly to all orders
by employing the renormalization group. Thus all necessary theoretical results
for calculating the Higgs signal and the non-Higgs continuum background contributions
to the process $\gamma \gamma \longrightarrow q \overline{q}$ are now known.
We are therefore able to present for the first time precise predictions for the
measurement of the partial width $\Gamma ( H \longrightarrow \gamma \gamma )$
at the Compton collider ($\gamma\gamma$) option at a future linear $e^+e^-$ collider.
The interplay between signal and background is very sensitive to the experimental
cuts and the ability of the detectors to identify $b$-quarks in the final state.
We investigate this in some detail  using a  Monte Carlo analysis, and conclude 
that a measurement with a 2 \% statistical accuracy
should be achievable. This could have
important consequences for the discovery of physics beyond the Standard Model, in particular
for large masses of a pseudoscalar Higgs boson as the decoupling
limit is difficult and for a wide range of $\tan \beta$ impossible 
to cover at the LHC proton-proton collider. 
\end{abstract}
\end{center}  
\vskip12pt

\setcounter{footnote}{0}  
\renewcommand{\thefootnote}{\arabic{footnote}}  
  
\vfill  
\clearpage  
\setcounter{page}{1}  
\pagestyle{plain} 
 
\section{Introduction} 
A central and so far experimentally unexplored element of the Standard Model 
(SM)
of particle physics is the origin of electroweak symmetry breaking. Indirect
evidence from precision measurements at $e^+e^-$ colliders suggests the existence
of a light Higgs boson in the mass range of $95 - 235$~GeV at the 95\% confidence
level with a statistical preference towards the lower end \cite{f}. 
{\sl A priori}, more complicated Higgs sectors are phenomenologically
just as viable. A well known example is provided by the general two doublet Higgs
model (2DHM) \cite{g,sh1,sh2}. 
A constrained version of the 2DHM is in fact a part of 
the minimal supersymmetric extension of the SM (MSSM), where spontaneous
symmetry breaking is induced by two complex Higgs doublets leading to five
physical scalars. The lightest of these is predicted to have a mass below the
$Z^0$ boson although radiative corrections can soften that limit up to about 
$130$~GeV due to the large value of the top mass.
 
In the MSSM there are only two parameters in the extended Higgs
sector, conventionally chosen to be the ratio of the vacuum expectation values of
the up and down type Higgs bosons, $\tan \beta$, and the mass $m_A$ of the predicted
neutral pseudoscalar Higgs. At tree level all other parameters are then fixed.

The main physics objectives of a future linear $e^+e^-$ collider depend crucially
on what
has been discovered at either LEP2, the Tevatron or the LHC. If no Higgs
bosons would be discovered at any of these machines 
the main objective will be to perform precision
tests of anomalous vector boson couplings to look for evidence of new physics 
at higher energy scales. If  supersymmetry is discovered, through the production
of new particles for example, then 
it is mandatory to investigate the precise structure of its manifestation in
order to hopefully have access to physics at scales where the Standard Model
couplings become equal and even gravity enters. Even if `only' a
SM Higgs is discovered, there will still be many unresolved issues 
concerning the validity domain and origin of SM physics. 

A common feature of all these possible scenarios is that a high degree of
experimental and theoretical precision will be required at a linear collider
to gain deeper insight into the structure of the physical laws of nature. 
In this context the photon-photon `Compton-collider' 
option, from backscattered laser light off
highly energetic and polarized electron beams \cite{g1,g2}, is a valuable ingredient.
It can provide complementary information about certain physical parameters which 
enter in different reactions, compared to the $e^+e^-$ mode, at comparable
event rates. In the Higgs sector, the Compton collider option offers a 
unique way to obtain  a precise determination of the partial
$\Gamma (H \longrightarrow \gamma \gamma)$ width. This quantity is important in
two respects. First, it allows for the model independent determination of the
total Higgs width, given that the appropriate branching ratio $BR(H \longrightarrow
\gamma \gamma)$ has been determined (at the LHC for instance).
Secondly, it is an important indicator of new physics as heavy charged 
particles which obtain
their mass through the Higgs  mechanism do {\sl not} decouple in the loop.
At the LHC, pseudoscalar masses above $m_A > 500$ GeV are not detectable for
a wide range of $\tan \beta$ and for intermediate values of $\tan \beta$ 
this regime reaches down to $\sim
250$ GeV \cite{s}.
In the MSSM this means that even in this so-called decoupling limit 
the diphoton partial Higgs width could
reveal the mass of the heavy pseudoscalar. Although the MSSM behaves rather
 like the SM for heavy $m_A$, studies suggest that the partial $\Gamma (H 
\longrightarrow \gamma \gamma)$ width can still differ by up to ten per cent
\cite{ddhi}.

It is clear from these lines of reasoning that it is very important at
this stage of the design of future linear colliders to know what the 
photon-photon option can contribute to the high precision measurements in the search
for  new physics. The purpose of this paper is  therefore to investigate 
the level of the statistical accuracy of the determination of the partial diphoton
Higgs width\footnote{It is clear that at this stage in the analyses we cannot
speculate about systematic errors.}. We focus on the processes $\gamma \gamma 
\longrightarrow q \overline{q}$  with $q=c,b$, and use all the available calculations 
relevant to these for both the intermediate mass Higgs signal as well
as the continuum  background. The Monte Carlo results are then combined
with expected Tesla machine and detector 
design parameters to arrive at reliable predictions for
the expected event rates. We begin by summarizing the status of the
QCD corrections to the tree-level processes. 

\section{Radiative Corrections to $\gamma \gamma \longrightarrow 
q \overline{q}$} \label{sec:rc}

In this section we begin by reviewing the QCD corrections to (continuum) heavy quark
production in polarized photon-photon collisions. As is well known, there are two possible
$J_z$ states, and at the Born level one has:
\begin{eqnarray}
\frac{d \sigma ( \gamma \gamma (J_z=0) \longrightarrow q \overline{q})}{d \; \cos
\; \theta} &=& \frac{12 \pi \alpha^2 Q^4_q \beta}{s \; ( 1 - \beta^2 \cos^2 \theta
)^2} (1-\beta^4)  \label{eq:BJ0} \\
\frac{d \sigma ( \gamma \gamma (J_z=\pm 2) \longrightarrow q \overline{q})}{d \; \cos
\; \theta} &=& \frac{12 \pi \alpha^2 Q^4_q \beta^3}{s \; ( 1 - \beta^2 \cos^2 \theta
)^2}(1- \cos^2 \theta)(2-\beta^2 (1-\cos^2 \theta)) \label{eq:BJ2}
\end{eqnarray}
where $\beta=\sqrt{1-4 m^2_q / s}$ denotes the quark velocity, $\sqrt{s}\equiv w$ 
the $\gamma\gamma$ c.m.
collision energy, $\alpha^{-1}
\approx 137$ the electromagnetic coupling, $Q_q$ the charge of quark $q$ and
$m_q$ its pole mass. The  scattering angle of the produced (anti)quark  
relative to the beam direction
is denoted by $\theta$. Eq.~\ref{eq:BJ0} clearly
displays the important feature that the $J_z=0$ cross section has a relative
$\frac{m^2_q}{s}$ suppression compared to  the $J_z=\pm 2$ cross
section \cite{inok,bar,gh,bkos}. 
Since the Higgs $\gamma\gamma \to H \to q  \overline{q}$
process only occurs for $J_z = 0$, it is this polarization that is crucial
for the precision measurement of the Higgs partial decay width.

However the overall $\frac{m^2_q}{s}$ factor in the $J_z=0$
cross section implies that, unlike for Eq.~\ref{eq:BJ2}, potentially large
radiative corrections
with logarithmic mass singularities are {\sl not} forbidden by the 
Kinoshita-Lee-Nauenberg
theorem \cite{k,ln}. This will become important in the next section
where we discuss large double logarithmic (DL) corrections to this cross section.
In contrast, the QCD corrections in the
$J_z=\pm 2$ case are not expected to be large.

For both helicity configurations the exact differential one-loop corrections are
known \cite{jt}
and are included in the calculations reported in this paper. 
For the $J_z=\pm 2$ helicity configuration, which, if we assume a high level
of efficiency for producing the $J_z = 0$ state, is expected to be heavily suppressed,
 these are sufficient for our purposes. For 
the zero helicity configuration the large logarithmic corrections {\sl are}
phenomenologically important, and so we discuss them in more detail
 in the next section.

\subsection{Double Logarithmic Form Factors}

The dominant background to Higgs production below 140~GeV is $\gamma \gamma (J_z=0)
\longrightarrow q \overline{q}$ with $q=b,c$. While this background is suppressed by
$\frac{m_q^2}{s}$ at the Born level, as shown above,
higher-order QCD radiative corrections in principle remove
this suppression \cite{bkos}. In addition, large virtual non-Sudakov double 
$\log(s/m_q^2)$ logarithms (DL)
are present which at one-loop can even lead to a negative cross section \cite{jt,jt2}.
The physical nature of these novel DL-effects was elucidated in Ref. \cite{fkm}.
There, the two-loop contribution to the non-Sudakov form factor was calculated
and it was shown that it allows for reasonable qualitative estimates. 
In particular,
positivity to the cross section was restored.
In Ref.~\cite{ms} the first explicit three-loop results in the DL
approximation were presented 
which revealed a factorization of non-Sudakov and
Sudakov double logarithms for this process and led to the all-orders
resummation in the form of a confluent hypergeometric function $_2F_2$: 
\begin{eqnarray}
\sigma^{DL}_{{\rm virt}+{\rm soft}} &=& \sigma_{\rm Born}
\left\{ 1 +
{\cal F} \;\;
_2F_2 (1,1;2,\frac{3}{2}; \frac{1}{2}
{\cal F} ) +
2 \; {\cal F} \;\;
_2F_2 (1,1;2,\frac{3}{2}; \frac{C_A}{4 C_F}
{\cal F} ) \right\}^2 \nonumber \\
&& \exp \left( \frac{ \alpha_s C_F}{\pi} \left[ \log \frac{s}{m_q^2} \left(
\frac{1}{2} - \log \frac{s}{4 l_c^2} \right) + \log \frac{s}{4 l_c^2} -1 +
\frac{\pi^2}{3} \right] \right) \label{eq:vps}
\end{eqnarray}
where ${\cal F} = - C_F \frac{\alpha_s}{4 \pi} \log^2 \frac{s}{m_q^2}$
is the one-loop non-Sudakov form factor,  $\alpha_s$ is taken as a fixed parameter,
and $l_c \ll \sqrt{s}$ is the soft-gluon upper energy limit.

In Ref.~\cite{ms2} it was pointed out that one needs to include at least 
{\it four} loops (at  the cross section level) of the non-Sudakov
logarithms in order
to achieve positivity and stability. At this level of approximation there is an
additional major source of uncertainty in
the scale choice of the QCD coupling, two possible `natural' choices ---
$\alpha_s(m_H^2)$ and $\alpha_s(m_q^2)$  --- 
yielding very different numerical results.  However in Ref.~\cite{ms3} 
this uncertainty was largely removed by
employing the renormalization group to introduce a {\sl running} QCD coupling.
We briefly summarize the results in the next section.

\subsection{Renormalization Group Improved Form Factors}

\begin{center}
\begin{figure}
\centering
\epsfig{file=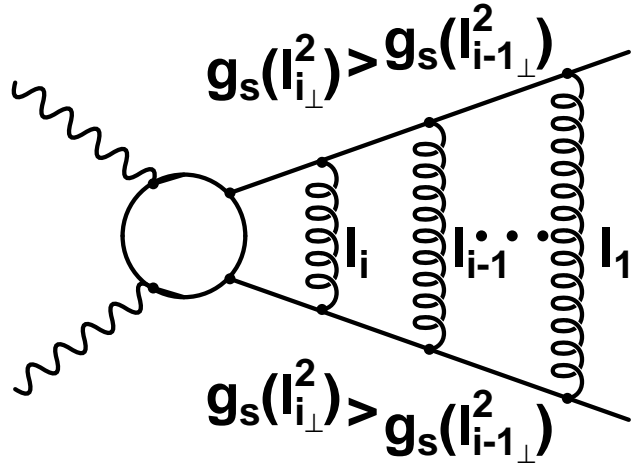,width=10cm}
\caption{A schematic Feynman diagram leading to the Sudakov
double logarithms in the process $\gamma \gamma (J_z=0) \longrightarrow q 
\overline{q}$ with $i$ gluon insertions. 
The blob denotes a hard momentum flowing through the omitted propagator
in the DL-phase space.
Crossed diagrams lead to a different
ordering of the Sudakov variables with all resulting $C_A$ terms canceling the
DL-contributions from three gluon insertions \cite{ms}.
The scale of the coupling $\alpha_s=\frac{g_s^2}{4
\pi}$ is indicated at the vertices and explicitly taken into 
account in this work.} 
\label{fig:SRG}
\end{figure}
\end{center}
In the derivation of the leading logarithmic corrections
in Ref.~\cite{ms} the familiar Sudakov technique \cite{Sud}
of decomposing loop momenta into components along 
external momenta, denoted by $\{\alpha, \beta\}$, and
those perpendicular to them, denoted by $l_\perp$ was used.  
For massless fermions, the effective scale for Sudakov double logarithms 
for the coupling at each loop
is $\alpha_s ( {\bf l^2_\perp} )$, as was shown e.g. in
Refs.~\cite{b,ddt,cl} by direct comparison with explicit higher-order
calculations. For massive fermions the effective scale is also given by
${\bf l^2_\perp} \equiv - l^2_\perp > 0$ as the dominant 
double logarithmic phase
space is given by $\frac{m^2}{s} \ll \frac{{\bf l^2_\perp}}{s} \ll 1$ \cite{ms}
(on a formal DL level, setting $m=\lambda$ yields the massless Sudakov 
form factor).
We use\footnote{This expression sums the leading order terms
($\sim \beta_0$) exactly.
The next-to-leading order terms ($\sim \beta_1$) are included correctly through
two-loops.}
\begin{equation}
\alpha_s ({\bf l^2_\perp})
 = \frac{\alpha_s(m^2)}{1+\beta_0 \frac{\alpha_s(m^2)
}{\pi} \log \frac{{\bf l^2_\perp}}{m^2}+\beta_1 \left( \frac{\alpha_s(m^2)}
{\pi} \right)^2 \log \frac{{\bf l^2_\perp}}{m^2}} 
\equiv \frac{\alpha_s(m^2)}{1+c \; \log \frac{{\bf l^2_\perp}}{m^2}} \label{eq:rc}
\end{equation}
where $\beta_0=\frac{11}{12} C_A -\frac{4}{12} T_F n_F$, 
$\beta_1 = \frac{17}{24} C_A^2 - \frac{5}{12} C_A T_F n_F- \frac{1}{4} 
C_F T_F n_F$ and for QCD we have
$C_A=3$, $C_F=\frac{4}{3}$ and $T_F=\frac{1}{2}$ as usual. 
Up to two-loops the massless $\beta$-function is independent of the chosen
renormalization scheme and is gauge invariant in minimally subtracted schemes
to all orders \cite{c}. These features also hold for the 
renormalization group improved form factors below.

For the massive Sudakov form factor we  use the on-shell condition 
${\bf l^2_{1_\perp}}=s \alpha_1 \beta_1$, even 
though the running coupling will now depend on {\it two} integration variables.
Since in this
work we are  only able  to include the hard {\it one-gluon} matrix elements
(the exact NNLO corrections are presently unknown),
the higher-order terms will inevitably  be energy cut ($l_c$) dependent. 
However our two-jet definition (see below) automatically restricts the higher-order 
hard gluon radiation phase
space such that it is reasonable to neglect more than one gluon emission.
The complete result for the renormalization
group improved massive Sudakov form factor is then given by \cite{ms3}:
\begin{eqnarray}
\widetilde{{\cal F}}^{RG}_{S_R}+2 \widetilde{{\cal F}}^{RG}_{S_V}&=& 
\frac{\alpha_s(m^2) C_F}{\pi} \left\{ \frac{1}{c}
\int_\frac{2l_cm^2}{(s+m^2) \sqrt{s}}^
\frac{2l_c\sqrt{s}}{s+m^2} \frac{d \beta_1}{\beta_1} \log \frac{ 1+c \; \log
\left( \left(\frac{2l_c}{\sqrt{s}} -\beta_1 \right) \beta_1 \frac{s}{m^2}
\right)}{\left(
1+c \; \log \frac{s \beta_1}{m^2} \right)} \right. \nonumber \\
&& - \frac{1}{c} \log \frac{s}{m^2} \log \frac{\alpha_s(2l_c \sqrt{s})}{
\alpha_s(s)} - \frac{1}{c}\log \frac{2l_c}{\sqrt{s}} \log \frac{\alpha_s (2 l_c \sqrt{s})}
{ \alpha_s \left( \frac{2l_cm^2}{\sqrt{s}}\right)} \nonumber \\ && \left. -
\frac{1}{c^2} \log \frac{\alpha_s(m^2)\alpha_s(2l_c \sqrt{s})}{
\alpha_s(s) \alpha_s \left(\frac{2l_c m^2}{\sqrt{s}} \right)} 
+ \frac{1}{2} \log \frac{s}{m^2} + \log \frac{s}{4 l_c^2} -1 + \frac{\pi^2}{3} \right\} 
\; ,
\label{eq:RpVRGex}
\end{eqnarray}
assuming only $\frac{m^2}{s} \ll 1$. Expanding in $\alpha_s(m^2)$ gives the
DL-Sudakov form factor in Eq.~\ref{eq:vps} together with  subleading terms proportional
to $\beta_0$ and subsubleading terms proportional to $\beta_1$\footnote{We
would like to point out that the Sudakov integration parameter $\beta_1$
entering into Eq. \ref{eq:RpVRGex} is not related to the two-loop
$\beta$-function coefficient.}. We emphasize
that the two-loop running coupling is included in Eq.~\ref{eq:RpVRGex} to
all orders and that all collinear divergences are avoided by
keeping all non-homogeneous fermion mass terms. In Ref.~\cite{ms3} it was shown
that Eq.~\ref{eq:RpVRGex} exponentiates in analogy to the soft exponential term
in Eq.~\ref{eq:vps}. The reason for adopting the above soft gluon energy
regulator rather than the more conventional $y_{cut}$ invariant mass
prescription (see Refs. \cite{bkos,jt,jt2} for instance) is connected with the
straightforward DL-phase space of the {\it massive} Sudakov form factor. The
latter is thus convenient for the inclusion of the renormalization group effects
as outlined above. More details are given in Ref. \cite{ms3}.
\begin{center}
\begin{figure}
\centering
\epsfig{file=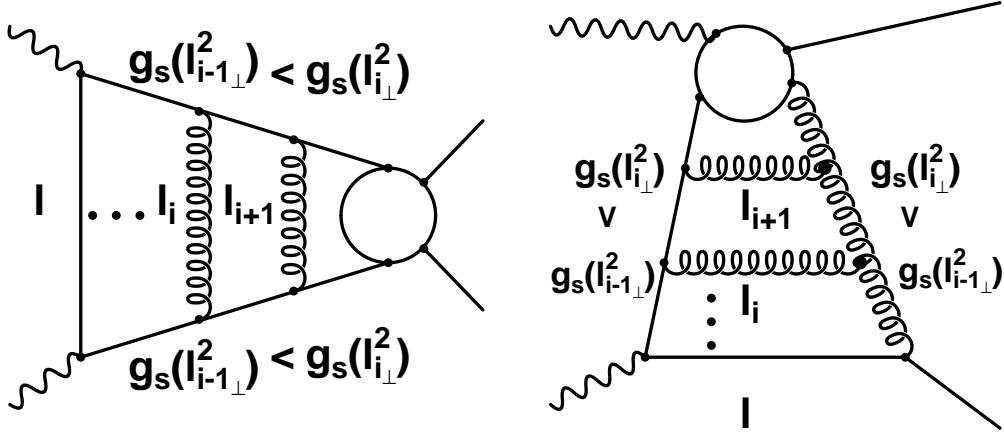,width=15cm}
\caption{The schematic Feynman diagrams leading to the non-Sudakov
double logarithms in the process $\gamma \gamma (J_z=0) \longrightarrow q 
\overline{q}$ with $i+1$ gluon insertions. 
The blobs denote a hard momentum flowing through the omitted propagator
in the DL phase space.
Crossed diagrams lead to a different
ordering of the Sudakov variables and are correctly accounted for by a factor
of $(i+1)!$ at each order. The scale of the coupling $\alpha_s=\frac{g_s^2}{4
\pi}$ is indicated at the vertices and included explicitly in the
calculation. The topology on 
the left-hand diagram is Abelian like,  and the one on the right is
non-Abelian beyond one-loop.}
\label{fig:RG}
\end{figure}
\end{center}
We next turn to the virtual non-Sudakov DL
corrections and investigate the RG effects for these
contributions.
Here we use the scale ${\bf l_\perp^2}$ directly as the graphs in Fig.~\ref{fig:RG}
are on a DL level identical to the Sudakov topology up to the last
integration over the (regulating) fermion line. This last integration, however,
does not renormalize the coupling.
For the non-Sudakov topologies depicted in Fig.~\ref{fig:RG} we find, 
after an order-by-order integration over the appropriate running coupling for
the complete
virtual renormalization group improved non-Sudakov form factor:
\begin{eqnarray}
\widetilde{{\cal F}}_h^{RG}
&=&\sum^\infty_{i=0} \int^s_{m^2} \frac{ d {\bf l^2_\perp}}{{\bf l^2_\perp}}
\left(\frac{C_F
}{2 \pi} \right)^{i+1} \left(
\frac{\alpha_s(m^2)}{ c}\right)^i
\frac{\alpha_s({\bf l^2_\perp})}{(i+1)!}
\log^{i+1} \frac{{\bf l^2_\perp}}{s} \log^i 
\frac{\alpha_s(m^2)}{\alpha_s ( {\bf l^2_\perp} )}
+  \nonumber \\
&&2 \sum^\infty_{i=0} \int^s_{m^2} \frac{ d {\bf l^2_\perp}}{{\bf l^2_\perp}}
\frac{C_F C_A^i
}{2^{2i+1} \pi^{i+1}} \left(
\frac{\alpha_s(m^2)}{c} \right)^i
\frac{\alpha_s({\bf l^2_\perp})}{(i+1)!}
\log^{i+1} \frac{{\bf l^2_\perp}}{s} \log^i 
\frac{\alpha_s(m^2)}{\alpha_s ( {\bf l^2_\perp} )} \; ,
\label{eq:rgff}
\end{eqnarray}
and thus for the RG-improved virtual plus soft real cross section we have
\begin{equation}
\frac{ \sigma^{DL}_{RG}}{\sigma_{\rm Born}} = \left\{ 1 + \widetilde{{\cal F}}_h^{RG}
\right\}^2 \exp \left( \widetilde{{\cal F}}^{RG}_{S_R} + 2 \widetilde{{\cal F}}^{RG}_{S_V}
\right) \label{eq:srg}
\end{equation}
\begin{center}
\begin{figure}
\centering
\epsfig{file=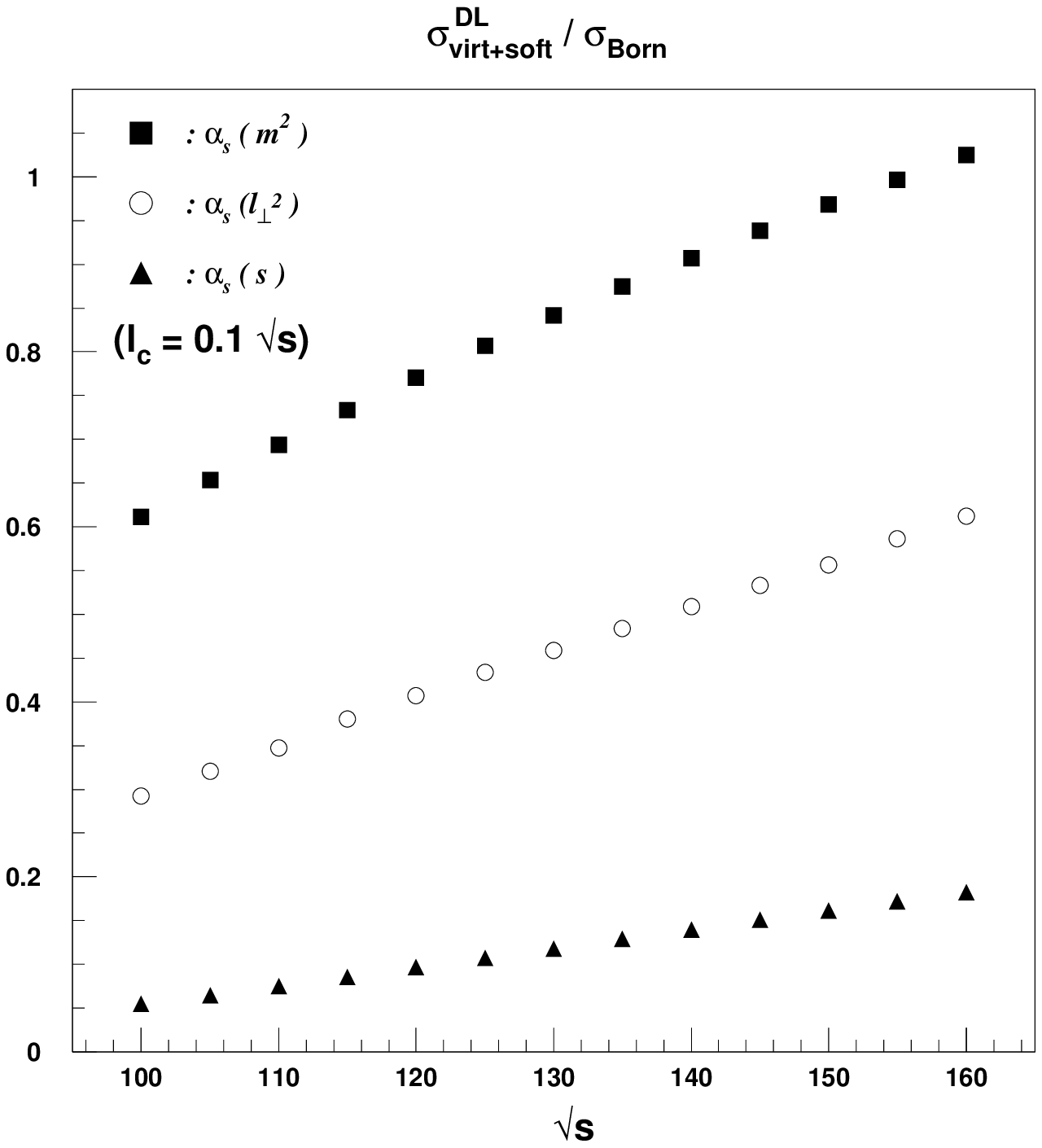,width=10cm} 
\vspace{-1cm} \\
\epsfig{file=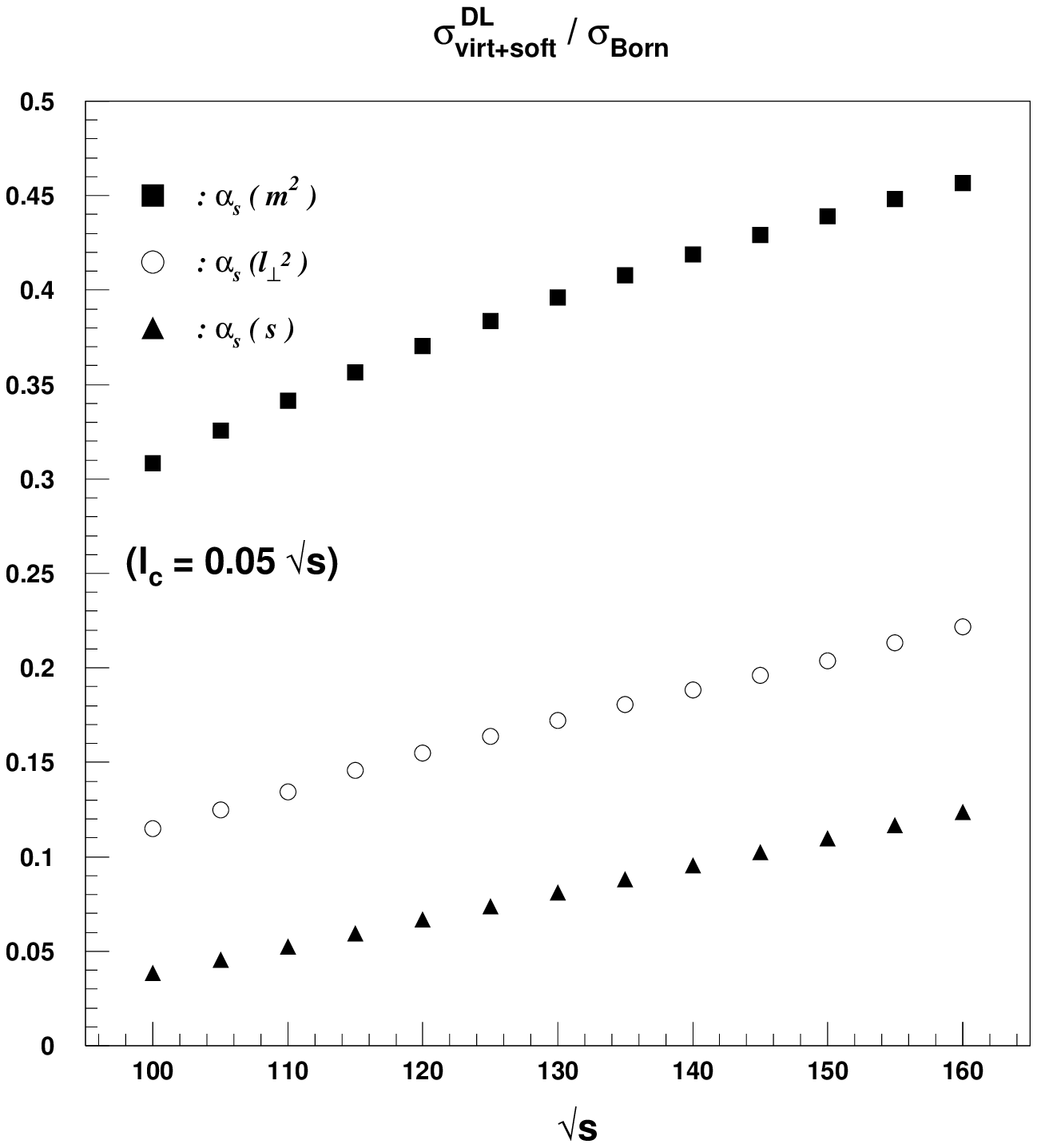,width=10cm}
\caption{The effect of the renormalization group improved form factor (circles)
of Eq.~\ref{eq:srg}
in comparison to using the DL form factors of Eq.~\ref{eq:vps} with the indicated
values of the strong coupling. The upper plot corresponds to $l_c=0.1 \sqrt{s}$ 
and the
lower one to $l_c=0.05 \sqrt{s}$. The effect is displayed for the bottom quark
with $m_b=4.5$ GeV.}
\label{fig:vps}
\end{figure}
\end{center}
where the RG-improved massive Sudakov form factor is given in Eq.~\ref{eq:RpVRGex}.
The effect of the renormalization group improved virtual plus soft real bremsstrahlung
cross section is depicted in Fig.~\ref{fig:vps}. The RG-improved 
cross section obtained 
using Eq.~\ref{eq:srg} lies  between the theoretically allowed upper 
and lower limits given by the double logarithmic form
factor of Eq.~\ref{eq:vps} with $\alpha_s$ evaluated at the bottom mass and
the Higgs mass scale. For a lower value of the energy cutoff $l_c$ the background
is more suppressed but the higher order (uncanceled) $l_c$-dependence is stronger.
This latter technical problem can be reduced by identifying $l_c$ with the
physical energy cutoff of the detector efficiencies. This will be discussed
in section \ref{sec:nr}. 

\section{Radiative Corrections to $\gamma \gamma \longrightarrow H \longrightarrow
q \overline{q}$} \label{sec:rcH}

An intermediate mass Higgs boson has a very narrow total decay width. It is therefore
appropriate to compare the total number of Higgs signal events with the number
of (continuum) background events integrated over a narrow energy window around
the Higgs mass. The size of this window depends on the level of monochromaticity
that can be achieved for the polarized photon beams.

\noindent In general, the number of events for the  (signal) process $S$
is given by
\begin{equation}
N_{S} = \int \frac{dL}{d w} \sigma_S(w) dw
\end{equation}
where $w$ denotes the center of mass energy. For $S \equiv \gamma \gamma \longrightarrow
H \longrightarrow b \overline{b}$ we have the following Breit-Wigner cross 
section, e.g. Refs. \cite{bbc1,bbc2}:
\begin{equation}
\sigma_S(w) = \frac{16 \pi \Gamma (H \longrightarrow \gamma \gamma) \Gamma ( H
\longrightarrow b \overline{b})}{(w^2-m^2_H)^2+\Gamma_H^2 m^2_H} (\hbar c)^2
\end{equation}
where the conversion factor $(\hbar c)^2=3.8937966\times 10^{11}$~fb~GeV$^2$.
In the narrow width approximation we then find for the expected number of 
events\footnote{In a realistic collider environment there will be a small
correction due to the fact that not 100\% of the incident photons are
polarized. These factors can easily be incorporated at a later time in parallel
with the exact luminosity distributions discussed below.}
\begin{equation}
N_S= \left. \frac{d L_{\gamma \gamma}}{dw} \right|_{m_H} \frac{8 \pi^2 \Gamma (
H \longrightarrow \gamma \gamma) BR ( H \longrightarrow b \overline{b})}{m^2_H}
(\hbar c)^2
\end{equation}
To quantify this, we take the design parameters of the proposed TESLA
linear collider  \cite{t,tp},
which correspond to an integrated peak
$\gamma \gamma$-luminosity of 15 fb$^{-1}$ for the low energy running of the
Compton collider. The polarizations of the incident electron beams and the
laser photons are chosen such that the product of the helicities $\lambda_e
\lambda_{\gamma} = -1$ \footnote{The maximal initial electron polarization
for existing projects is 85 \%, e.g. Ref. \cite{t}.}.
This ensures high monochromaticity and polarization of the photon beams \cite{g1,g2,t,tp,gnk}.
Within this scenario a typical resolution of the Higgs mass is about 10~GeV, so 
that for comparison with
the background process $BG \equiv \gamma \gamma \longrightarrow q \overline{q}$
one can use \cite{bbc1,bbc2}:
\begin{equation}
\frac{L_{\gamma \gamma}}{10\; {\rm GeV}} = \left. \frac{d L_{\gamma \gamma}}{dw} 
\right|_{m_H}
\end{equation}
with $\left. \frac{d L_{\gamma \gamma}}{dw} \right|_{m_H}=$0.5 fb$^{-1}$/GeV.
The number of background events is then given by
\begin{equation}
N_{BG} = L_{\gamma \gamma} \sigma_{BG}\; .
\end{equation}
In other words,
 the number of signal events is proportional to $N_S \sim \left. \frac{d 
L_{\gamma \gamma}}{dw} \right|_{m_H}$ while the number of  continuum heavy
quark production events is proportional to $N_{BG} \sim L_{\gamma \gamma}$.
In principle it is possible to use the exact Compton profile of the backscattered
photons to obtain the full luminosity distributions. The number of expected
events is then given as a convolution of the energy dependent luminosity and
the cross sections. Our approach described above corresponds to an effective
description of these convolutions, since these functions are not precisely
known at present. Note that the functional forms currently used
generally assume that only one scattering takes place for each photon, which may not
be realistic.  Once the exact luminosity functions
are experimentally determined  it is of course trivial to
incorporate them into a Monte Carlo program containing the physics described
in this paper.
\begin{center}
\begin{figure}[t]
\centering
\epsfig{file=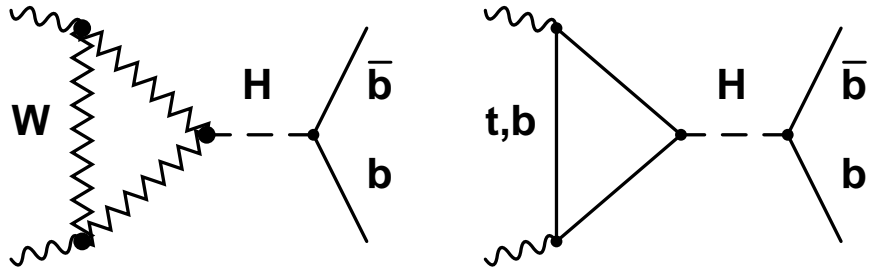,width=12cm}
\caption{The Standard Model process $\gamma \gamma \longrightarrow H \longrightarrow
b \overline{b}$ is mediated by  $W-$boson and  $t$- and $b$-quark loops.}
\label{fig:H}
\end{figure}
\end{center}
We next summarize the radiative corrections entering into the 
calculation of the expected number of
Higgs events.
For the quantity $\Gamma (H \longrightarrow \gamma \gamma)$ there are three main
Standard Model contributions, depicted in Fig. \ref{fig:H}: the $W^\pm$
and $t$- and $b$-quark loops. We include these at the one-loop level, since the 
radiative corrections are significant  only for
the $b$-quark  \cite{s}. 
The branching
ratio $BR(H \longrightarrow b \overline{b})$ is treated in the following way.
The first component consists of the partial $\Gamma (H \longrightarrow b \overline{b}
)$ width. Obviously we must use the same two-jet criterion 
for the signal as for the background. For our purposes a cone-type algorithm is most
suitable, and so we use the Sterman-Weinberg two-jet definition \cite{sw}
depicted in   Fig.~\ref{fig:jet}. Note that the signal cross section is corrected
by the {\sl same} resummed renormalization group improved form factor
given in Eq.~\ref{eq:RpVRGex}, since this factor does not depend on the spin of the
particle coupling to the final state quark anti-quark pair.
\begin{center}
\begin{figure}[t]
\centering
\epsfig{file=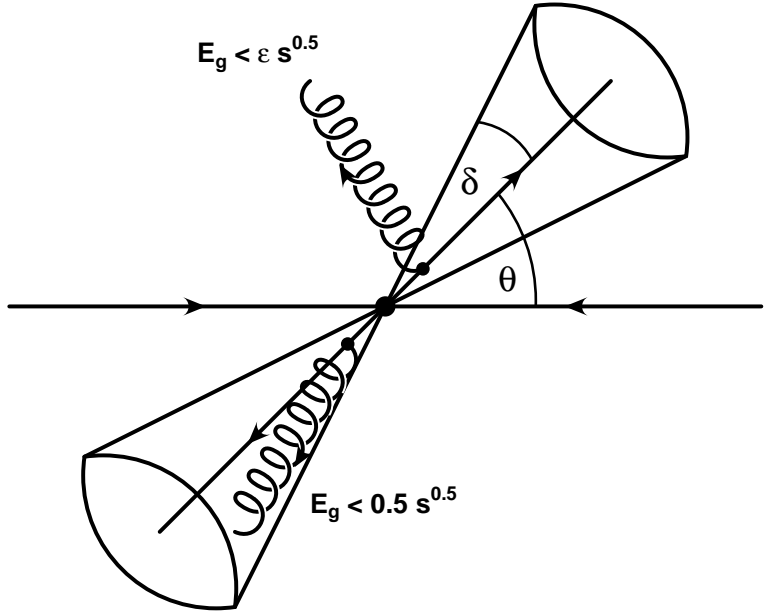,width=10cm}
\caption{The parameters of the Sterman-Weinberg two-jet definition used in this
work. Inside an angular cone of size $\delta$ arbitrary hard gluon bremsstrahlung
is included. Radiation outside this cone is only permitted if the gluon energy
is below a certain fraction ($\epsilon$) of the incident center of mass energy. 
The thrust angle is denoted by $\theta$.}
\label{fig:jet}
\end{figure}
\end{center}

In addition we use the exact one-loop corrections from Ref.~\cite{bl}. These
revealed that
the largest radiative corrections are well described by using the running quark
mass evaluated at the Higgs mass scale. We therefore resum the leading running
mass terms to all orders. 
For the real bremsstrahlung corrections we use our own $H \longrightarrow 
q \bar q g$ 
matrix elements. An important 
check is obtained by integrating over all phase space and reproducing the analytical
results of Ref.~\cite{bl}. 

The second quantity entering the branching ratio is the total Higgs width $\Gamma
_H$. Here we use the known results summarized in Ref.~\cite{s}, and include
the partial Higgs to $b \overline{b}$, $c \overline{c}$, $\tau^+\tau^-$, $WW^*$, 
$ZZ^*$ and $gg$ decay widths with all relevant radiative corrections.

\section{Numerical Results} \label{sec:nr}

In Ref.~\cite{ms3} numerical predictions were given for an (infra-red safe)
two-jet $b \bar b$ cross section in $\gamma\gamma$ 
collisions in the energy
range $\sqrt{s} = 100  - 160$~GeV. A modified 
Sterman-Weinberg cone definition, imposed on the final state partons was 
employed.
 Thus, at leading order
(i.e. $\gamma\gamma \to b \overline{b}$) all events obviously
satisfy the two--$b$--jet
requirement.\footnote{We apply an angular cut of
$\vert{\cos\theta_{b,\overline{b}}}\vert
< c_0$ to ensure that both jets lie in the central region of the detector.} 
This defines
our `leading order' (LO) cross section. 
At next--to--leading order (NLO) we can have virtual or real gluon emission.
For the latter, an event is defined as two--$b$--jet like if the emitted
gluon 
\begin{eqnarray*}
\mbox{{\it either}}&& \mbox{I.\quad ~has energy less than $\epsilon \sqrt{s}$, with  
$\epsilon \ll 1$}, \\
\mbox{{\it or}} && \mbox{II.\quad is within an angle 
$2 \delta$ of the $b$ or $\overline{b}$, again
with $\delta \ll 1$}.
\end{eqnarray*}
 We further subdivided 
region I according to whether
the gluon energy is greater or less than the infrared cutoff $l_c$
 ($< \epsilon$). Adding the virtual gluon corrections to this latter
(soft)  contribution, to give $\sigma_{\rm SV}$, and calling the remaining
hard gluon contribution $\sigma_{\rm H}$, we have 
\begin{equation}
\sigma_{\rm 2j} = \sigma_{\rm SV}(l_c) + 
\sigma_{\rm H}(l_c,\epsilon,\delta) \; .
\label{eq:2jet}
\end{equation}
In Ref.~\cite{ms2} each part of this cross section was evaluated exactly
to ${\cal O}(\alpha_s)$ and in addition the resummed
non-Sudakov form factor was included in $\sigma_{\rm SV}$. This was necessary
to yield a positive cross section.

\noindent We use the RG-improved expressions
for the resummed form factors. Thus 
\begin{equation}
\sigma_{\rm SV } \longrightarrow \sigma^{DL}_{RG} + 
\tilde\sigma_{\rm SV }\; ,
\end{equation}
where $\sigma^{DL}_{RG}$ is given in Eq.~\ref{eq:srg} and 
$\tilde\sigma_{\rm SV }  $ is the exact one-loop result minus the one-loop 
leading-logarithm pieces which are resummed in $ \sigma^{DL}_{RG}$, i.e.
\begin{eqnarray}
\tilde\sigma_{\rm SV }  &=& \sigma_{\rm SV, NLO} - 
\sigma_{\rm LO} \left[ -6 \cF  +  \frac{ \alpha_s C_F}{\pi} 
\left( \log \frac{s}{m_q^2} \left(
\frac{1}{2} - \log \frac{s}{4 l_c^2} \right) + \log \frac{s}{4 l_c^2} -1 +
\frac{\pi^2}{3} \right) \right] \; . \nonumber \\
&&
\end{eqnarray}
By adding the second (Sudakov) piece in the square brackets we remove
(at least up to terms $\cO(l_c^2/s) \ll 1$) the dependence on the gluon
energy cutoff $l_c$. 
Note also that the complete expression for the two-jet cross section 
(with the remaining $l_c$ dependence displayed)
\begin{equation}
\sigma_{\rm 2j} = \sigma^{DL}_{RG}(l_c) + \tilde\sigma_{\rm SV} + 
\sigma_{\rm H}(l_c,\epsilon,\delta) \; .
\label{eq:2jetbis}
\end{equation}
contains  a mixture of exact ${\cal O}(\alpha_s)$ and resummed pieces.
For the former, we use $m_q^2$ as the scale for $\alpha_s$.\footnote{We choose
the QCD scale parameter $\Lambda$ such that $\alpha_s(m_b^2)=0.2235$
for $m_b = 4.5$~GeV, at both leading and next-to-leading
order.}
The resummed contributions are based on the 
scale choice ${\bf l^2_\perp}$ in the loops, as already discussed.

Before computing  and combining the various components of the two-jet cross section
in Eq.~\ref{eq:2jetbis} we must address the issue of the dependence on the unphysical
infra-red parameter $l_c$. If we were to expand out the resummed RG-improved form factor
$\sigma^{DL}_{RG}(l_c)$ in powers of $\alpha_s(m_q^2)$, and retain only
the $\cO(\alpha_S)$ term, we would find that the $l_c$ dependence exactly canceled
that of $\sigma_{\rm H}(l_c,\epsilon,\delta)$.\footnote{This was shown explicitly
in Ref.~\cite{ms2}, see for example Fig.~3 therein.} However in the full resummed expression,
there is nothing to cancel the explicit $l_c$ dependence at higher-orders. The canceling
terms would come from the as yet unknown higher-order contributions to $\sigma_{\rm H}$.
Faced with this dilemma, we have several choices. We could, as in \cite{ms2}, neglect
the higher-order terms in the 
Sudakov form factor altogether, and include only the non-Sudakov form factor
which is of course independent of $l_c$. Furthermore, as shown in \cite{ms2}
with the choice $\epsilon=\cO(0.1)$, the combined 
contribution of virtual gluons and real gluons with $E_g < \epsilon \sqrt{s}$
to $\sigma_{\rm 2j}$ was dominated by the non-Sudakov `$6 \cF$' part. This 
suggests that the most reasonable procedure for the resummed
cross section is to take $l_c \sim \epsilon \sqrt{s}$ and to vary
$\epsilon$. We stress that this 
is an {\it approximation}, since it corresponds to making an assumption
about the contribution of real multi-gluon emission with energies 
$ < \cO(\epsilon\sqrt{s})$.

As our `best guess' RG-improved, resummed two-jet cross section, therefore, we have
\begin{equation}
\sigma_{\rm 2j} = \sigma^{DL}_{RG}(\epsilon\sqrt{s}) + \tilde\sigma_{\rm SV} + 
\sigma_{\rm H}(\epsilon\sqrt{s},\epsilon,\delta) \; .
\label{eq:2jetbisbis}
\end{equation}

At this point it is appropriate to comment on detector and accelerator 
related issues which were adopted
in our analyses \cite{ba}.
Since we are not using a full detector simulation we 
employ effective performance parameters which should be achievable at a future
linear collider. We will display results for realistic scenarios for both
currently accepted and more optimistic cases. In particular the double b-tagging efficiency
will be assumed to be $70\%$ throughout and the main input parameters concern
the probability of counting a $c \overline{c}$-pair as $b \overline{b}$ and the
ratio of the photon-pairs in a $J_z=0$ to $J_z=\pm2$ state. We emphasize again
that these dependences are in real machine environment given by 
functional forms which can
easily be determined through test runs at a later stage. For our purposes here the
effective description is sufficient. 

The results discussed in the next section
contain all radiative corrections summarized above. The goal is to
optimize the jet-parameters of the Sterman-Weinberg
two-jet definition in order to maximize the statistical significance of the
intermediate mass Higgs-boson signal.

\subsection{Discussion of the MC Results}

We begin with a few generic remarks concerning the uncertainties in our
predictions. The signal process $\gamma \gamma \longrightarrow H \longrightarrow
b \overline{b}$ is well understood and NNL calculations are available. The 
theoretical error is thus negligible \cite{s}. 

There are two contributions to 
the background process
$\gamma \gamma \longrightarrow q \overline{q}$ which we neglect in this paper.
Firstly, the so-called resolved photon contribution
\cite{dkzz} was found to be a small effect, e.g. \cite{jt,jt2}, especially since
we want to reconstruct the Higgs mass from the final two-jet measurements and
impose angular cuts in the forward region. In addition the good charm suppression
also helps to suppress the resolved photon effects as they give the
largest contribution.
The second contribution we do not consider here results from the final state configuration where
a soft quark is propagating down the beam pipe and the gluon and remaining
quark form two hard back-to back-jets \cite{bkos}. We neglect this contribution
here due to the expected excellent double b-tagging efficiency and the strong 
restrictions on the
allowed acollinearity discussed below \footnote{As discussed in Ref. \cite{bkos}
the B-hadrons from the slow b-quark could be dragged towards the gluon side and
thus give rise to displaced decay vertices in the gluon jet. It may be of interest
to perform further systematic MC studies of this effect.}.

For the continuum heavy quark production cross section
an exact NLO calculation exists
\cite{jt} but large radiative corrections in the $J_z=0$ channel require the
resummation of large non-Sudakov DL's as expounded on above. Assuming that the
largest part of the NLO and higher subleading logarithms is contained in the
renormalization of the strong coupling parameter $\alpha_s$, the virtual
corrections seem to be well under control. The largest uncertainty we thus expect
to be contained in the missing hard ${\cal O} \left( \alpha_s^2 \right)$ 
bremsstrahlung corrections for which no $\frac{m_q^2}{s}$ suppression-factor
exists. Theoretically, these can be controlled by limiting the available
phase-space through a narrow two-jet definition. On the other hand this means
that we would also lose (signal) events which is clearly undesirable. 
In light of these
two effects we think it prudent to find a balancing middle ground for our
MC-results\footnote{The precise size of the background process can be
determined by scanning the energy regions below and above the Higgs resonance.
The exact functional form is still necessary to obtain a precision measurement
of $\Gamma (H \longrightarrow \gamma \gamma)$ for resonant energies.}. More details are given below.

A second source of uncertainty is contained in the higher order $l_c$-dependence
as mentioned above. Our strategy of identifying $\epsilon \sqrt{s} = l_c$ is
reasonable as long as the neglected ${\cal O} \left( \epsilon^2 \right)$ terms
(which have no Born cross section suppression) are negligible and can be
identified with the physical detector energy cuts. In this paper we will thus
study two different values for the energy cut: $\epsilon = 0.1$ and $0.05$. 
The value of $\epsilon$ is related to the allowed acollinearity of the two
jet alignments corresponding to acollinearities of $11.5^o$ and $5.73^o$.
We emphasize that the requirement of a small jet acollinearity
substantially suppresses the $J_z=\pm2$ background component and could play
an important role in improving the photon collider energy resolution
\cite{t,tp} as well as in the suppression of the background due to the
resolved processes. 
Below we display results assuming in each case a (realistic) ratio
of $J_0/J_2=20$ in parallel with the (optimistic) ratio of $J_0/J_2=50$.

\begin{center}
\begin{figure}
\centering
\epsfig{file=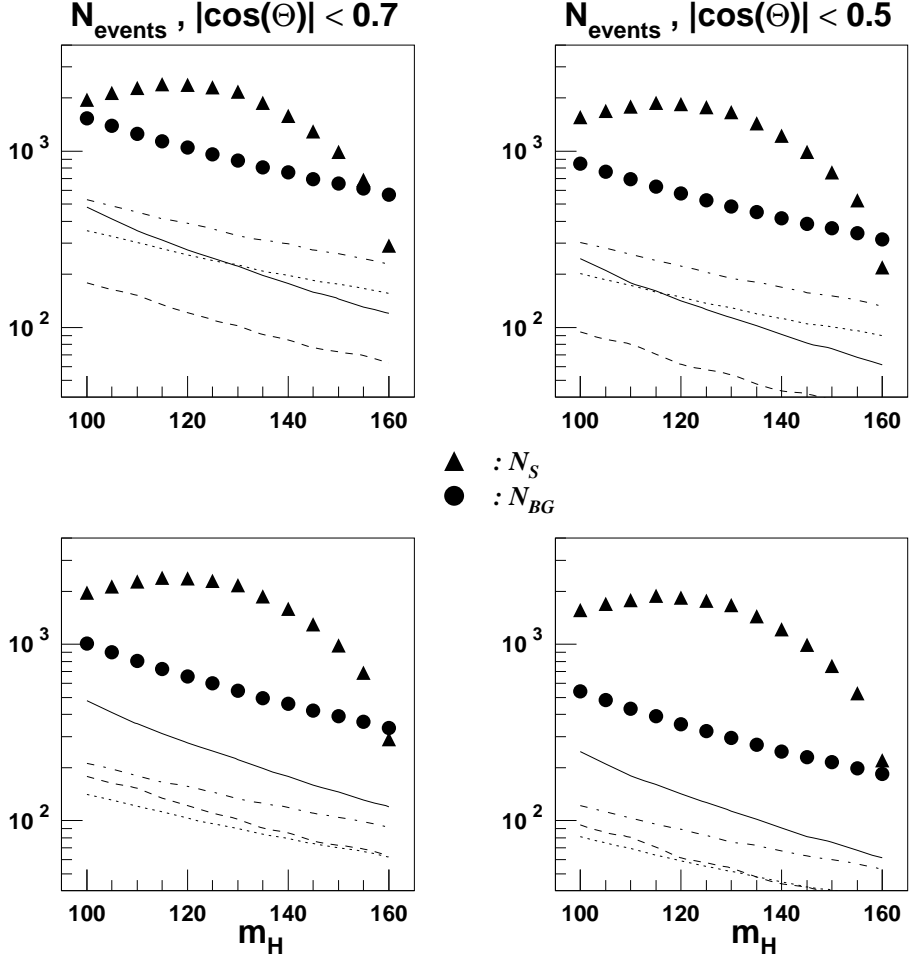,width=15cm}
\caption{The number of both signal and background events for jet
parameters $\epsilon=0.1$ and $\delta=20^o$ and the indicated values
of the thrust angle $\theta$. The upper row assumes a ratio of
$J_0/J_2=20$ and the lower row of $50$. The background is composed of
bottom and charm contributions assuming 70 \% double b-tagging efficiency
and a 3 \% probability to count a $c \overline{c}$ pair as $b \overline{b}$.
The dash-dotted line corresponds to $J_z=\pm2$ for $m_c$, the full line to
$J_z=0$ for $m_b$, the dotted line to $J_z=\pm2$ for $m_b$ and the dashed
line to $J_z=0$ for $m_c$. All lines are are normalized to add up to the
total background and all radiative corrections discussed in the
text are included.}
\label{fig:ne203}
\end{figure}
\end{center}

We start with Fig. \ref{fig:ne203} assuming a (quite realistic) probability
of counting a $\cc$- as a $\bb$-pair of $3\%$ and the Sterman-Weinberg
parameters $\epsilon=0.1$ and $\delta=20^o$. The figure shows signal and
BG events separately for two values of the thrust angle $\theta$ cut, $| \cos
\theta | < 0.7$ and $| \cos \theta | < 0.5$. In both scenarios it can be seen that
the largest component to the BG events for $J_0/J_2=20$ 
originates from the $J_z=\pm2$ $c$-quark
contribution. The second largest corrections stem from the $b$-quark for both
$J_z=0$ and $J_z=\pm2$ while the $J_z=0$ $c$-quark contribution is small.

\begin{center}
\begin{figure}
\centering
\epsfig{file=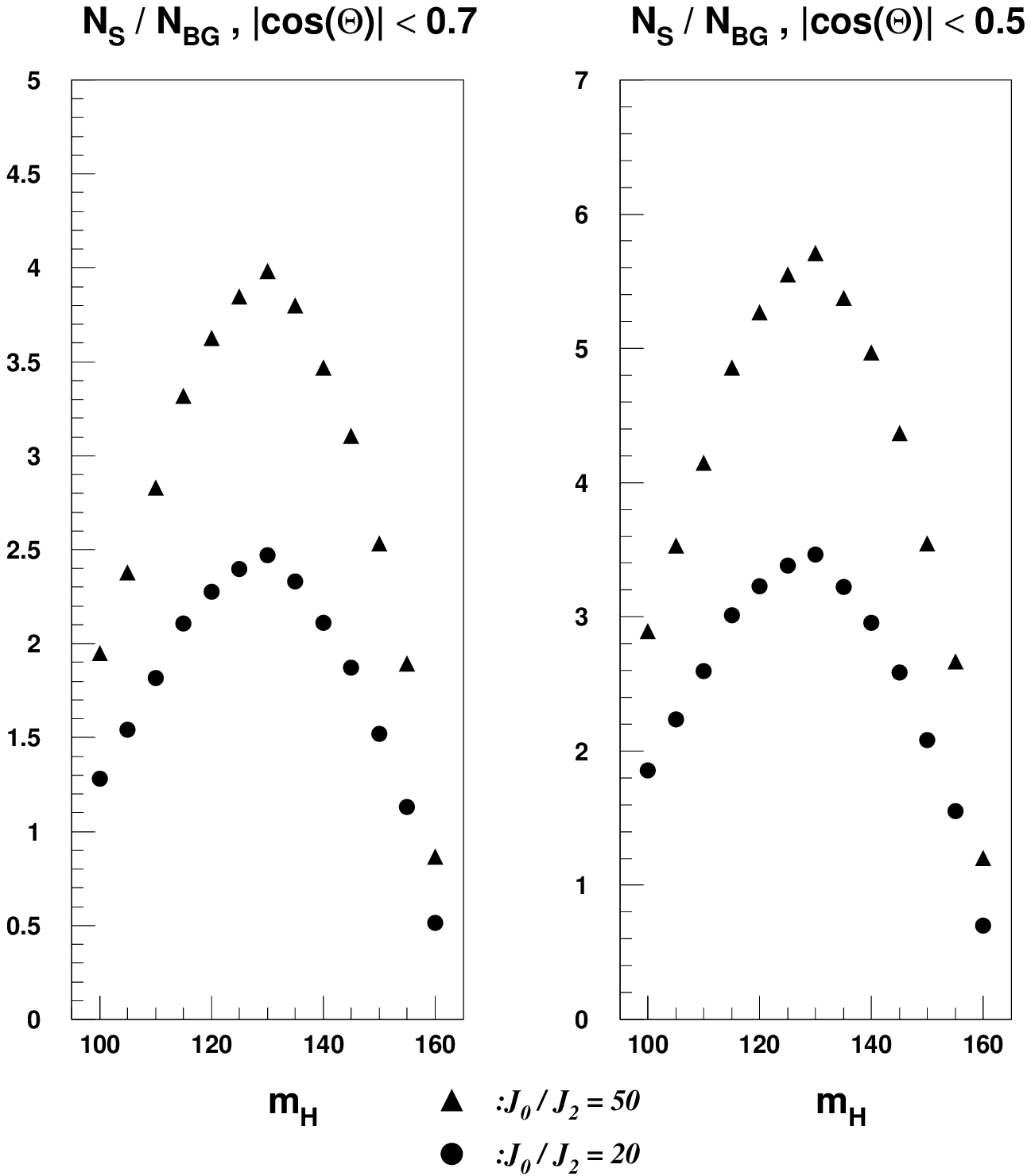,width=15cm}
\caption{The ratio of signal to background events based on the jet parameters
of Fig. \ref{fig:ne203}. The smaller phase space cut $| \cos \theta |<0.5$ gives
a larger ratio as expected.}
\label{fig:sbg203}
\end{figure}
\end{center}

\begin{center}
\begin{figure}
\centering
\epsfig{file=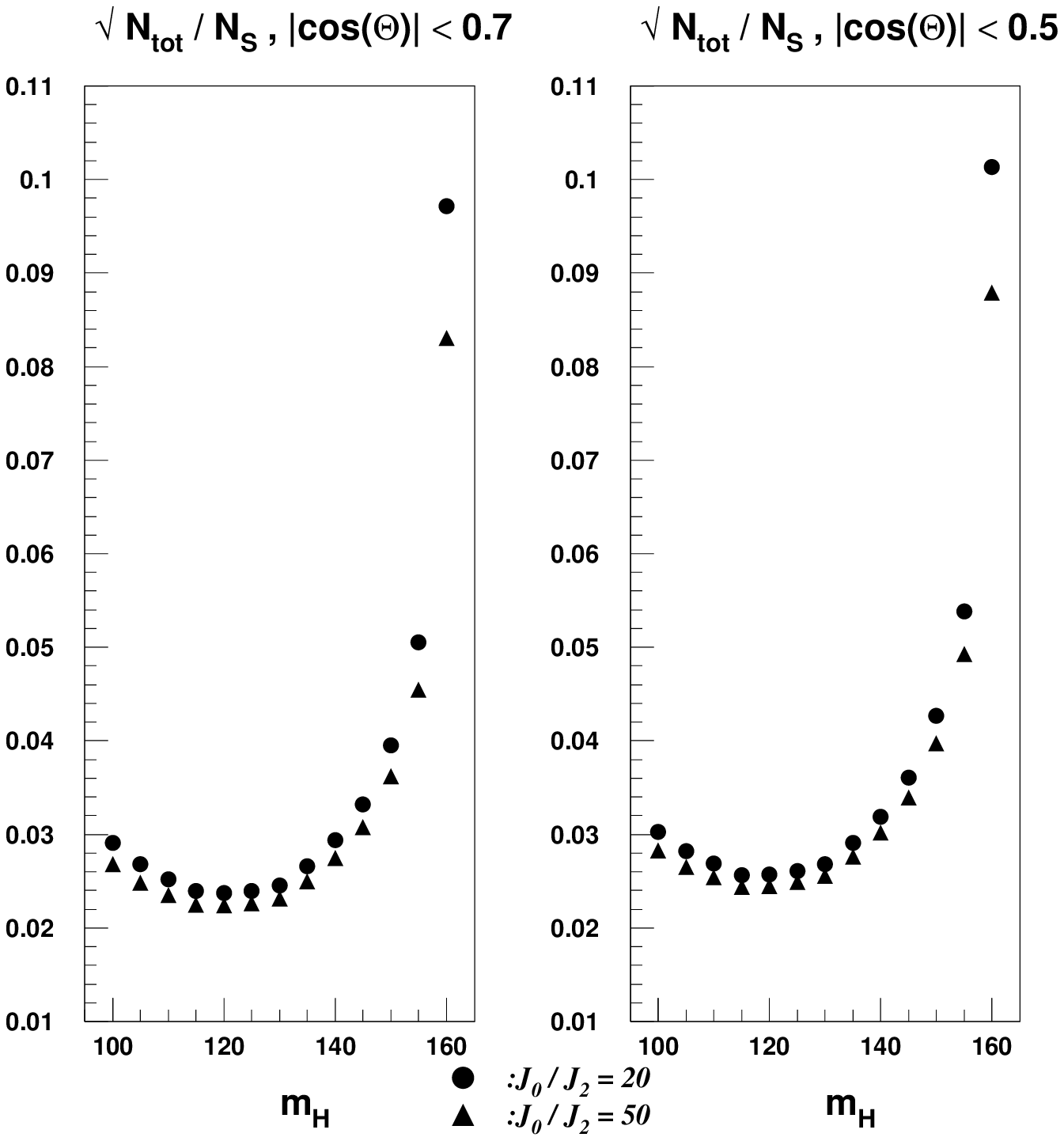,width=15cm}
\caption{The statistical accuracy of the measurement based on a one year running
with the parameters of Fig. \ref{fig:ne203}. The larger thrust angle
cut gives a slightly better statistical significance.}
\label{fig:sa203}
\end{figure}
\end{center}

The smaller thrust-cut two-jet definition eliminates more of the background
events in relative terms. However, it also reduces the total number of events.
Fig. \ref{fig:sbg203} demonstrates that the $| \cos \theta | < 0.5$ scenario
yields roughly a 50 \% higher ratio of signal to BG events. The inverse 
statistical
significance of the Higgs-boson process, defined as $\sqrt{N_{tot}} / N_S$, however,
is somewhat higher for the $| \cos \theta | < 0.7$ choice as is demonstrated in Fig.
\ref{fig:sa203}. The difference between the realistic $J_0/J_2=20$ 
and optimistic 
$J_0/J_2=50$ photon-polarization cases is small. For the one-year 
running analysis of an intermediate mass Higgs with $m_H < 140$GeV the
inverse statistical significance is below 3\%, which can be viewed as the minimal
statistical expectation.

\begin{center}
\begin{figure}
\centering
\epsfig{file=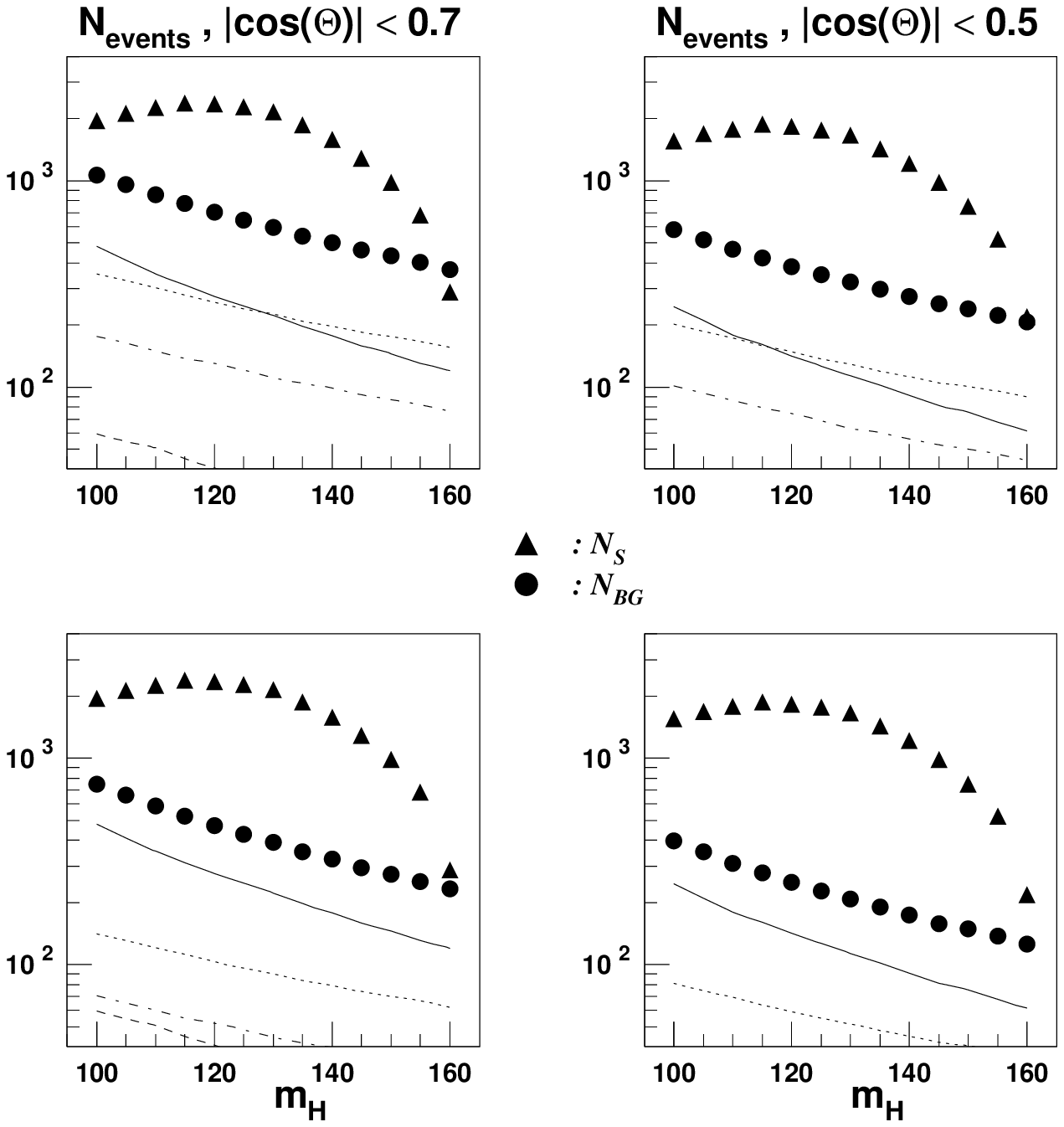,width=15cm}
\caption{The number of both signal and background events for jet
parameters $\epsilon=0.1$ and $\delta=20^o$ and the indicated values
of the thrust angle $\theta$. The upper row assumes a ratio of
$J_0/J_2=20$ and the lower row of $50$. The background is composed of
bottom and charm contributions assuming 70 \% double b-tagging efficiency
and a 1 \% probability to count a $c \overline{c}$ pair as $b \overline{b}$.
The dash-dotted line corresponds to $J_z=\pm2$ for $m_c$, the full line to
$J_z=0$ for $m_b$, the dotted line to $J_z=\pm2$ for $m_b$ and the dashed
line to $J_z=0$ for $m_c$. All lines are are normalized to add up to the
total background and all radiative corrections discussed in the
text are included.}
\label{fig:ne201}
\end{figure}
\end{center}

\begin{center}
\begin{figure}
\centering
\epsfig{file=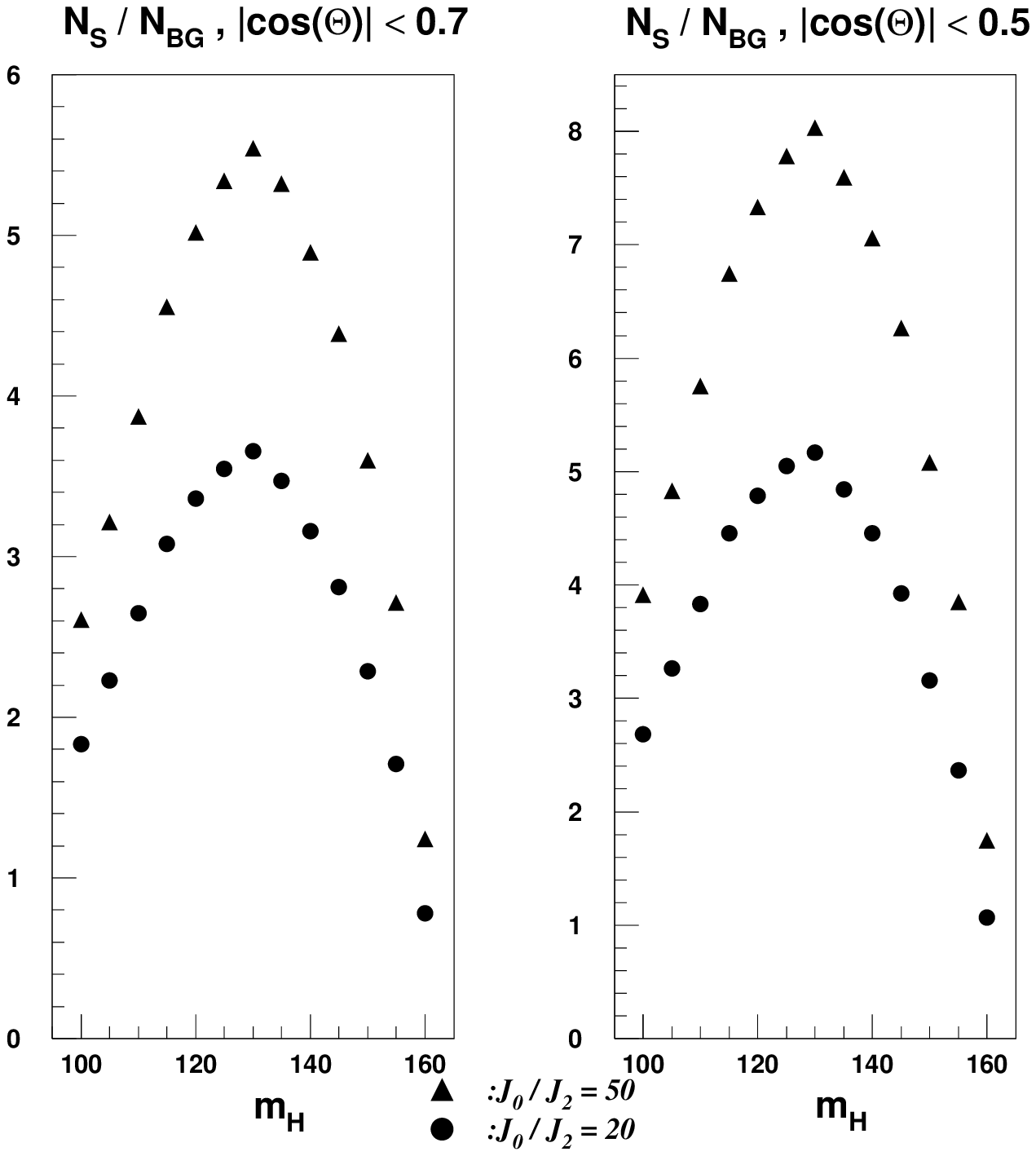,width=15cm}
\caption{The ratio of signal to background events based on the jet parameters
of Fig. \ref{fig:ne201}. The smaller phase space cut $| \cos \theta |<0.5$ gives
a larger ratio as expected.}
\label{fig:sbg201}
\end{figure}
\end{center}

\begin{center}
\begin{figure}
\centering
\epsfig{file=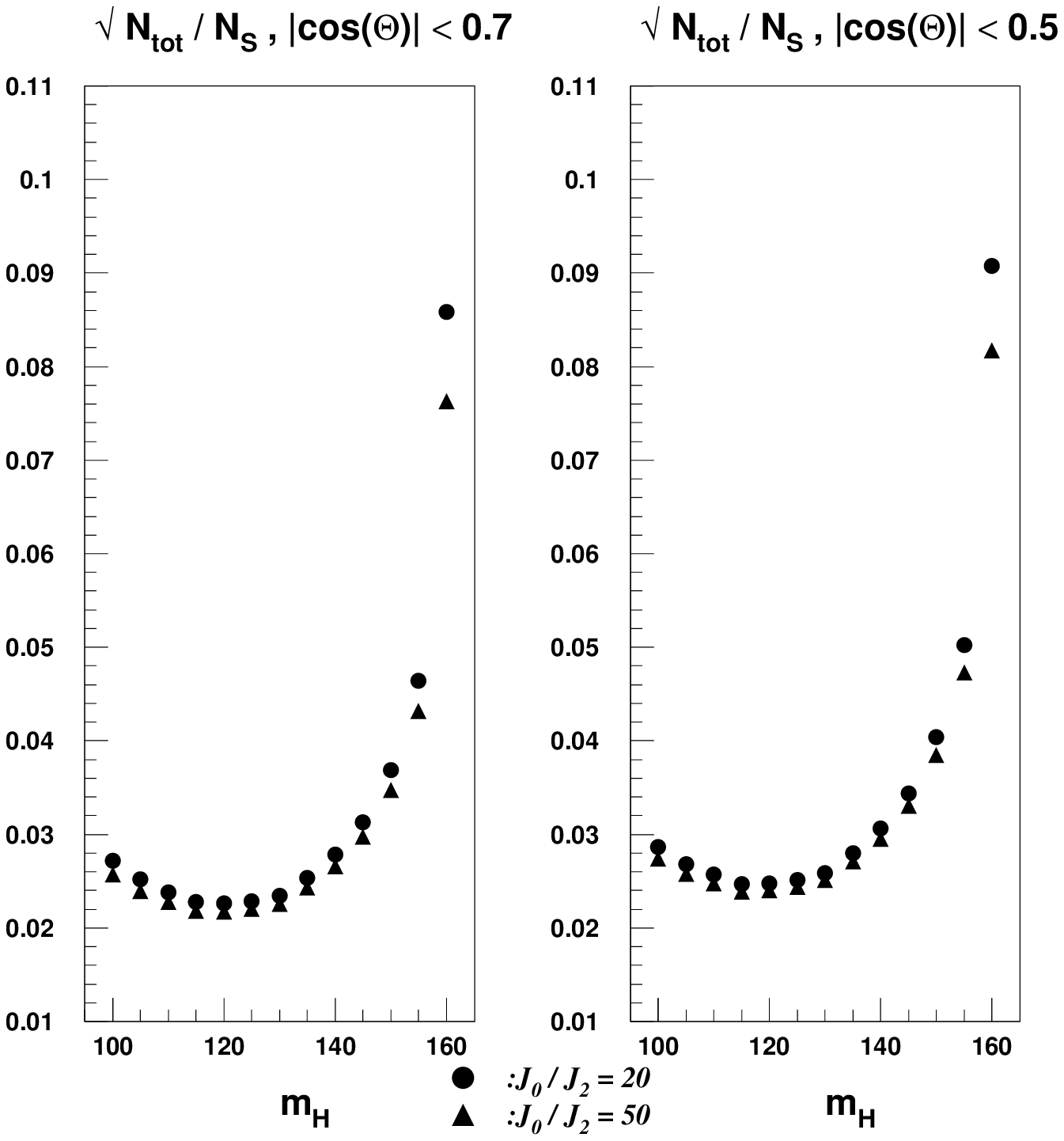,width=15cm}
\caption{The statistical accuracy of the measurement based on a one year running
with the parameters of Fig. \ref{fig:ne201}. The larger thrust angle
cut gives a slightly better statistical significance.}
\label{fig:sa201}
\end{figure}
\end{center}

It seems now possible to assume an even better detector performance. 
The improvement comes
from assuming a better single point resolution, thinner detector modules and 
moving the vertex detectors closer to the beam-line \cite{ba}.
Thus we can assume a (still realistic) 1\% probability of counting
a $\cc$- as a $\bb$-pair. Figs. \ref{fig:ne201}, \ref{fig:sbg201}
and \ref{fig:sa201} display
the same observables for otherwise identical two-jet definitions and
machine-parameters. The charm-contribution is visibly reduced and the
number of signal to background events roughly 30\% larger. The statistical
accuracy for the Higgs signal, however, is only slightly enhanced.

\begin{center}
\begin{figure}
\centering
\epsfig{file=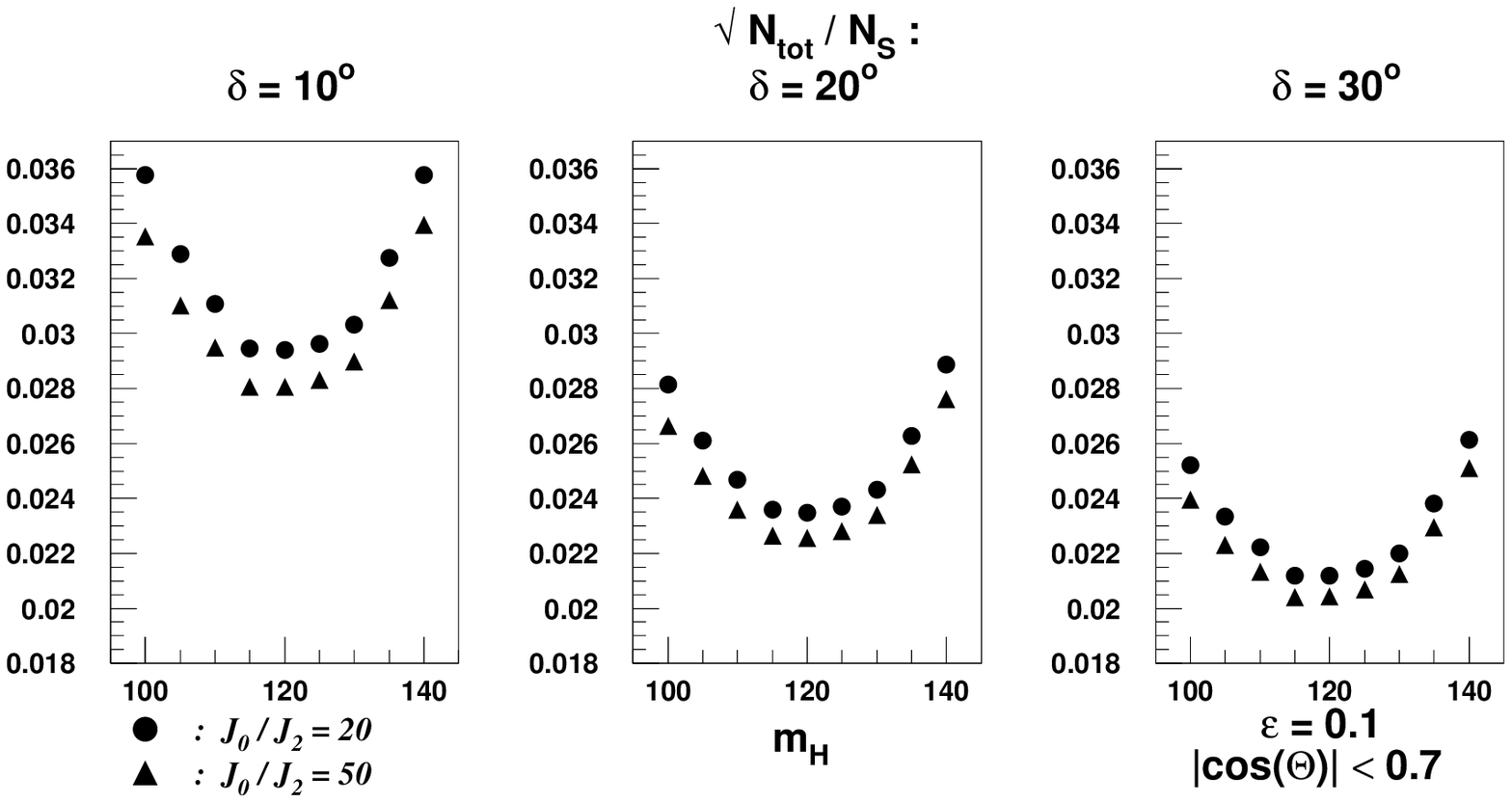,width=17cm}
\epsfig{file=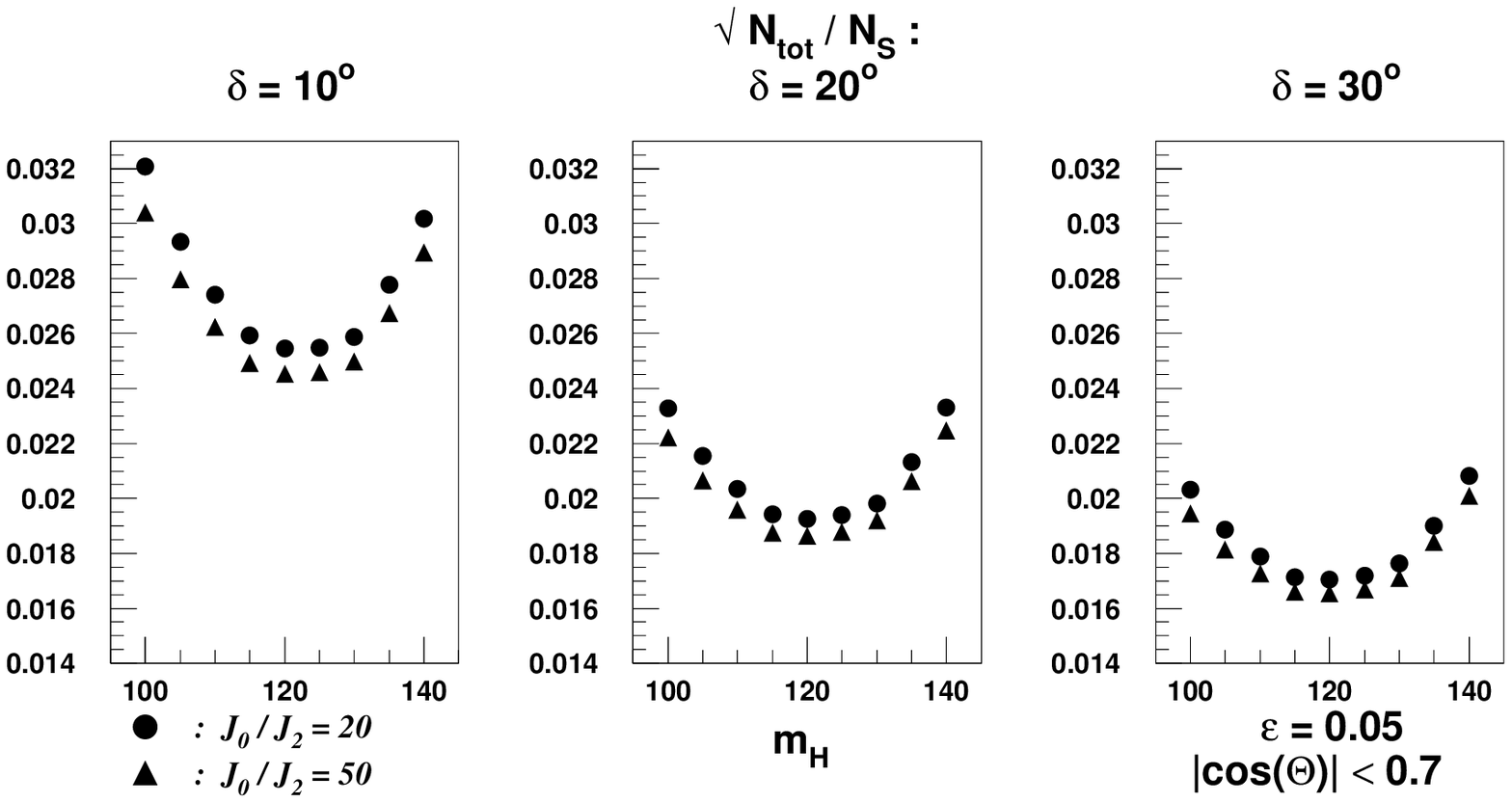,width=17cm}
\caption{The cone-angle dependence of the inverse statistical significance
of the intermediate mass Higgs signal for the displayed values of thrust
and energy cut parameters. Overall a 70\% double b-tagging efficiency and
a 1\% charm misidentification rate are assumed. For larger values of $\delta$
the number of events is enlarged, however, the theoretical uncertainty increases.
For smaller values of $\epsilon$ higher order cutoff dependent terms might become
important.}
\label{fig:cs}
\end{figure}
\end{center}

With these results in hand we now keep $| \cos \theta | < 0.7$ fixed
and furthermore assume the $\cc$ misidentification rate of 1\%. We vary
the cone angle $\delta$ between narrow ($10^o$), medium ($20^o$) and large
($30^o$) cone sizes for both $\epsilon=0.1$ and $\epsilon=0.05$.
The upper row of Fig. \ref{fig:cs} demonstrates that for the former 
choice of the energy
cutoff parameter we achieve the highest statistical accuracy for the
large $\delta=30^o$ scenario of around 2\%. We again emphasize, however, that
in this case also the missing ${\cal O} \left( \alpha_s^2 \right)$ bremsstrahlung
corrections could become important.

The largest effect is obtained by effectively suppressing the background
radiative events with the smaller energy cutoff of $\epsilon=0.05$ outside
the cone (the inside is of course independent of $\epsilon$). Here the
lower row of Fig.
\ref{fig:cs} demonstrates that the statistical accuracy of the Higgs
boson with $m_H < 140$ GeV can be below the 2\% level after collecting one
year of data. We should mention again that for this choice of $\epsilon$
we might have slightly enhanced the higher order (uncanceled) cutoff dependence.
The dependence on the photon-photon polarization degree is visible but not
crucial.

In summary, it seems very reasonable to expect that at the Compton collider
option we can achieve a 2\% statistical accuracy of an intermediate mass
Higgs boson signal after collecting data over one year of running.

\section{Conclusions} \label{sec:con}

In this paper we have studied the Higgs signal and continuum background
contributions to the process $\gamma\gamma \longrightarrow
b \overline{b}$ at a high-energy Compton collider.
We have used all relevant QCD
radiative corrections to both the signal and BG production
available in the literature. The Monte Carlo results using a variety of
jet-parameter variations revealed that the intermediate mass 
Higgs signal can be expected to be studied with a statistical uncertainty
between $3\%$ in a very realistic and $1.6\%$ in an optimistic scenario after
one year of collecting data.

Together with the expected uncertainty of 1\% from the $e^+e^-$ mode determination
of BR$(H \longrightarrow \bb)$, and assuming four years of collecting data, we
conclude that statistically a measurement of the partial width
$\Gamma (H \longrightarrow \gamma \gamma)$ below the 2\% precision 
level should be possible.
This level of accuracy could significantly enhance the kinematical reach
of the MSSM parameter space 
in the large pseudoscalar mass limit and thus open up a window for
physics beyond the Standard Model. 
 
For the total Higgs width, the main uncertainty is given by the error in the
branching ratio BR$(H \longrightarrow \gamma \gamma)$, which at present is
estimated at the 15 \% level \cite{brient}. For Higgs masses above 110 GeV,
the total Higgs width could be determined more precisely through the
Higgs-strahlung process \cite{egn,ik} and its decay into $WW^*$ \cite{br}. 
This is only possible, however, if the supersymmetric lightest Higgs boson
coupling to vector bosons is universal (i.e. the same for $hWW$ and $hZZ$) 
and provided the
optimistic luminosity assumptions can be reached.

In summary, using conservative machine and detector design parameters, we 
conclude that the Compton collider option at a future linear collider
can considerably extend our ability to discriminate between the SM and
MSSM scenarios.

\vspace{1cm}

\noindent{\bf Acknowledgements}\\ 
We would like to thank M.~Battaglia, G.~Jikia, R.~Orava, M.~Piccolo and
especially V.I.~Telnov for useful discussions.
This work was supported in part by the EU Fourth Framework 
Programme `Training and Mobility of 
Researchers', Network `Quantum Chromodynamics and the Deep 
Structure of Elementary Particles', 
contract FMRX-CT98-0194 (DG 12 - MIHT).  

\end{document}